# Case study: Data Mining of Associate Degree Accepted Candidates by Modular Method

**Behrouz Minaei Bidgoli; Maryam Nazaridoust**

*Department of Computer Engineering and Information Technology, University of Qom, Qom, Iran*

*\*nazaridoustm@gmail.com*

*b_minaei@iust.ac.ir*

## Abstract

Since about 10 years ago, University of Applied Science and Technology (UAST) in Iran has admitted students in discontinuous associate degree by modular method, so that almost 100,000 students are accepted every year. Although the first aim of holding such courses was to improve scientific and skill level of employees, over time a considerable group of unemployed people have been interested to participate in these courses. According to this fact, in this paper, we mine and analyze a sample data of accepted candidates in modular 2008 and 2009 courses by using unsupervised and supervised learning paradigms.

In the first step, by using unsupervised paradigm, we grouped (clustered) set of modular accepted candidates based on their student status and labeled data sets by three classes so that each class somehow shows educational and student status of modular accepted candidates. In the second step, by using supervised and unsupervised algorithms, we generated predicting models in 2008 data sets. Then, by making a comparison between performances of generated models, we selected predicting model of association rules through which some rules were extracted. Finally, this model is executed for Test set which includes accepted candidates of next course then by evaluation of results, the percentage of correctness and confidentiality of obtained results can be viewed.

## 1. Introduction

Higher education is always faced with huge data and information about universities, students, scholars, professors, and other resources; so that in most of cases, this data can include valuable information and patterns. Using modern techniques of data mining such as predicting, clustering, classification, etc can be useful in ranking of universities, finding specific and valuable patterns about successful students, finding a successful teaching programs or method, finding critical points in financial management of university, and so on. Also, discoverable knowledge through data mining in accepted candidates can be used and we can offer advices to participants to have better performance and make better decisions.

Almost a decade has been passed from holding modular courses of UAST University and has yielded plenty of graduates so far and although the first aim of holding such courses was to improve scientific and skill level of employees, over time a considerable group of unemployed people have been interested to participate in these courses. So that up to now, more than 100,000 people are admitted in modular associate degree courses and even because of high interest of applicants, modular undergraduate (bachelor) courses have been hold recently. In this admission system, The National Education Assessment admits students generally based on their Grade, employment history, the relevancy of their job with requested field, sexuality, and kind of diploma (high school degree) and accepted students should pass comprehensive exam to get related degree documents.

In the first section of paper, data sets and data initialization method will be introduced briefly. In the second section, we will have a glance on model generating by using unsupervised learning and supervised learning. In the third section, by a representation on clustering and its kinds, two suitable clustering techniques will be selected and then executed on data sets through which data can be labeled based on generated clusters. In the fourth section, two predicting algorithms are considered for data sets, and then performances of the algorithms will be compared. In the fifth section, by using selected predicting algorithm, two separate models will be introduced for data sets. First model presents some rules in order to predict associate degree fields for new candidates. Second model, by another perspective, according to student conditions of candidate, extracts the rules which predict the only student status class for new candidate. Ultimately, in the last section, conclusions of performed analysis will be presented.

## 2. Data Sets

Data set used in this study is a sample of accepted candidates in modular associate degree in 2008 and 2009 which contains about 20,000 of accepted candidates during two years. Features of this data set are described in Table1.

**Table 1.  Features of this data set**

| | Attribute Name | Attribute Value |
|---|---|---|
| 1 | ID | -- |
| 2 | Gender | Female,Male |
| 3 | Grade(Average)[1] | 12< |
| 4 | Age | 17< |
| 5 | Kind of diploma | high school degree |
| 6 | Employment history | Unemployed, employed |
| 7 | Relevancy of  job | Unemployed=0 , Related employed=1 , Unrelated employed=2 |
| 8 | Field Group | Industry , Management and Social Services, Culture and Art, Agricultural |
| 9 | Field | IT,Software, management and …. |

### 2.1. Data Preparing

Data preparing is the most important and time consuming step in data mining projects and as the output strongly is dependent on input, more precise input can be resulted in more precise and reliable output. Otherwise, we will be confronted with GIGO (Garbage in Garbage Out) phenomenon in outputs.

Pre process step of data includes data transformation from main resource and its conversion to reasonable and suitable structure for interested data mining algorithms. Therefore, before using data mining algorithms, pre process applications should be performed in order to initialize data.[2] In brief, performed tasks are making data integrated and uniform, outlier and missing value data evaluation, reducing dimensions and generating new specification. For instance, an available specification in initial data set was birth year of accepted candidates which was converted to age based on date for better using.

---

[1] This grade is between 0 -20

[2] In this study, pre process has been done by using two application programs (Clementine and Data mining tab in Microsoft Office Excel 2007)

## 3. Generating Models

Nowadays by Information Technology development, understanding and realizing high volume of data have found more importance; therefore, solutions and practices have been more difficult and complicated. In order to get better knowledge about huge databases and their analysis, some models can be generated. One of the models is an application program or intelligent process which, through using data history, detects the patterns and important procedures that have major and essential impact on business criteria. Then, such detected patterns and procedures can be used to forecast possible results which makes us able to include taking suitable actions in calculations and affect on future results potentially. In other words, a model is an automated process which discovers patterns and procedures in order to generate rules. Each model can be generated based on supervised and unsupervised learning algorithms. Unsupervised learning detects available structure in data by an automated process with no specified label. The supervised paradigm is unlike the previous case, which knowledge detecting process in a set of data is done with supervising; that is such learning intends to mention and classify some areas of a specific context.

## 4. Clustering

Clustering has been raised as a conceptive format and rich algorithm to analyze and interpret data and is followed by organizing and discovery of structure in collected data sets. In most cases, clustering is considered as a synonym of "unsupervised learning" so that generally in clustering, without considering predetermined classes, heterogeneous population of dataset is divided to some homogenous clusters.

In this method, data is grouped merely based on existing similarities and differences (e.g. distance) among data points and final clusters should have two features: 1) high homogenous inside each cluster, and 2) heterogeneous among different clusters. Even, we can interpret aim of clustering as generating some classes and assignment of data to generated data. Lots of strategies have been proposed for clusters generating which have caused some kinds of clustering algorithms having different and diverse features. Clustering techniques can be divided into 3 groups: (1) Segmentation based clustering or function based, (2) hierarchical clustering, and (3) model based clustering.

### 4.1 Suitable clustering selection

Grouping of similar data is completely subjective and strongly is dependent upon clustering criterion. Based on this issue we can argue that so far varied clustering algorithms have been generated which offer relatively different results. Figure 1 shows result of applying three different clustering on a unique data set.

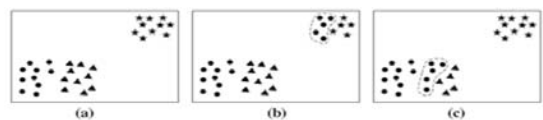

**Figure 1. Three clustering, each containing three clusters. a Predefined clustering; b clustering P; c clustering S**(resource :[7])

According to variety of clustering algorithms and different used methods, the first question made for us was "which clustering algorithm would be chosen by us?"

Researchers usually use such indices as Rand index, Jacard index, and even ADCO (Attribute Distribution Clustering Orthogonally) to compare and measure similarity between two clustering.

Rand and Jacard indices act through pair counting, their comparisons are based on membership and philosophy is based on differentiation. Such indices are compatible with user's aim of clustering where differentiation of objects should be performed. But, ADCO scale considers another method and is parallel with the philosophy based on predicting model for clustering. Eric Bae et al (2008) evaluated algorithms of every three clustering by considering criteria and as a result announced that K-means differentiation clustering algorithms has had better performance for under survey data sets. According to this result, we also used K-means clustering algorithm to group accepted candidates community of this course.

## 4.2. Generating a dividing model by using two clustering algorithms

As shown in previous section, based on results yielded by indices, K-Means clustering algorithms has revealed better performance for clustering; but for optimal execution on K-Means clustering algorithm, we need to know K suitable number for clusters. Therefore, initially by using hierarchical clustering method, optimal number of clusters can be obtained and then by performing K-Means clustering application, we will specify members of each cluster:

## 4.3. Two Step Algorithm

This algorithm is one of the hierarchical methods which use distance criterion based on Log_Liklihood possibility. To calculate this criterion, numerical variables need to have normal distribution and non-numerical discrete variables need to follow polynomial distribution. One of the specifications of TwoStep algorithm is its capability to determine optimal number of clusters automatically.

By performing TwoStep algorithm on student specifications of accepted candidates, optimal number of clusters is obtained equal 3. Result of this algorithm execution is shown in Figure 2.

## 4.4. K_Means Algorithm

This algorithm is one of the segmentation based methods which use variance minimum criteria to organize data. This algorithm is one of the simplest unsupervised learning algorithms which need predetermined number of K to group data, so the main idea of this algorithm is to define K centers for each cluster. Such centers should be selected carefully, because different centers cause different results. Therefore, further distance of these centers with each other, better result. Next step is to assign each pattern to the closest center. Once all the points are assigned to existing centers, an initial grouping has been done; therefore, new K centers need to be calculated for previous clusters and former process should be repeated. This task will be repeated until having no displacement to another K center.

In this study, by relying on key specifications which seem to be more effective to evaluate accepted candidates, we performed K-Means algorithm on data set and considered clusters number (K) of 3 based on result yielded from TwoStep hierarchical clustering. Result of this algorithm implementation is shown in Figure 3.

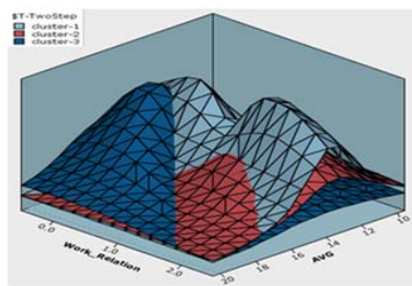

**Figure 2. Cluster Scatter diagram with two feature : of K- Means Relevancy of job**

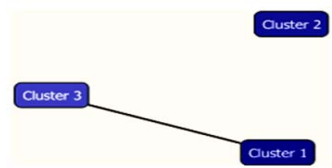

**Figure 3.  The result and Means clustering**

## 4.5. Surveying specifications of clusters and labeling cluster's items

According to K-means clustering results and former figure, it is absolutely clear that clusters 2 and 3 are different from many perspectives and clusters 1 and 3 have more similarities with each other. Generally by this method, we could separate candidates' community to three clusters based on their student statues (high school) and label each cluster with one class that according to above mentioned specifications, such three classes have the most possible difference and distance with each other.

**Cluster 1**: About 72% of persons in this cluster are 25 to 31 years old and more than 90% of them are men and have an Grade in a range of 13 to 15, most of them are unemployed or have unrelated job with their selected field.

**Cluster 2**: 78% of the population in this cluster are 17 to 25 years old, so this cluster is the youngest one whose majority (more than 74%) is possessed by ladies; and more than half of the candidates in this cluster have an Grade more than 15 and most of them are unemployed.

**Cluster 3**: About 77% of the candidates in this cluster are 31 to 59 years old; most of them are employee and have requested a field compatible with their occupation, and 68% of them have an Grade in range of 12 to 15.

In the fifth section, by using a predicting model, labeled datasets will be trained and through such model, status of applicants for next courses will be predicted in terms of their student status.

## 5. Generating predict model

In this step, we are looking for a predict model which can be used to predict admission of next courses candidates. To do such a work, we can choose different patterns among which decision tree algorithms and dependency rules algorithm are taken into account which are of supervised and unsupervised algorithms, respectively.

### 5.1. Decision tree model

One of the most conventional rating algorithms is decision tree which is one of the main parts of modern machines learning and is put in category of supervised learning algorithm. The main aim in decision tree is to divide data recursively to some subsets so that each subset can include a homogenous status of target variable. This algorithm does some predictions based on relations among input columns in a dataset and uses values and status of such columns to predict target value status (as predictable column). A decision tree generates the decision rules that each of them is a realizable conditional phrase which is used readily inside the database to detect a set of records.

One of the advantages of decision tree is to generate precise and interpretable models with relatively low interference of user. This algorithm can be used for binary and multiclass rating problems whose rules and scoring, in visual perspective, is done in parallel sort.

### 5.2. Association Rules

Association Rules extraction is one of the unsupervised and so important methods of data mining. By using such a technique, we can find interesting relations and dependencies in data set. Often, interesting and useful rules discovery provides an information resource through which the performance can be better and more suitable decisions can be made. Association Rules can be shown as following:

LHS—RHS [Support, Confidence]

To measure quality of a rule, <u>support</u> and <u>confidence</u> are used. Support declares which number of member samples of dataset has been used to generate rule which includes cases related to LHS and RHS (together). Confidence specifies which number of items including cases related to LHS also contains cases related to RHS. If an association rule can present some minimum values, it would be more suitable. Extraction of Association Rules is done when datasets are real descriptors which happen simultaneously (or close to each other).

### 5.3. Generating predict models and results comparison

Initially, in order to select the best predict model, we generate two models by using both algorithms that yielded dependency network by both models has been shown in figures 4 and 5. These two figures show performance difference between both algorithms.

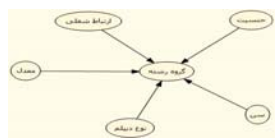

**Figure 4. Above figure shows dependency Network of Decision Three**

**Table 2. Some rules of first predicting model**

| Rule# | Gender | & Grade | & diploma | & Relevancy of job | Field(RHS) |
|-------|--------|---------|-----------|--------------------|------------|
| 1 | F | 14.8 - 16.3 | Math-Physics | 0 | Software |
| 2 | M | >12.7 | Job and Knowledge | 2 | Car Quality Control & Machine Tools |
| 3 | M | <12.7 | Technical & Professional | 1 | Financial Services in Trade Units |
| 4 | F | 14.8 −16.3 | Art | 0 | Graphic |
| 5 | M | <12.68 | Human Sciences | 0 | Accounting- Industrial |
| 6 | M | - | Job and Knowledge | 2 | Software |

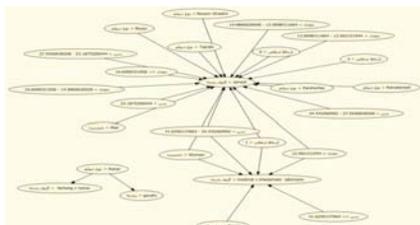

**Figure 5.  Opposite figure shows dependency network of Association Rules**

Of course, by observing rise procedure of lift chart, it may be imagined that these two algorithms have no sensible difference in terms of measurement criteria and model yielded by decision tree has had relatively better performance. Even, scores inserted in Figure 4 emphasize on relatively better performance of yielded model by decision tree model in comparison with Association Rules model. However, it should be noted that privilege value of such a model is negligible. But, why we chose latter model to predict admission possibility of candidates in next course? We believe that, according to yielded total information by Figure 6, Association Rules model had had better performance.

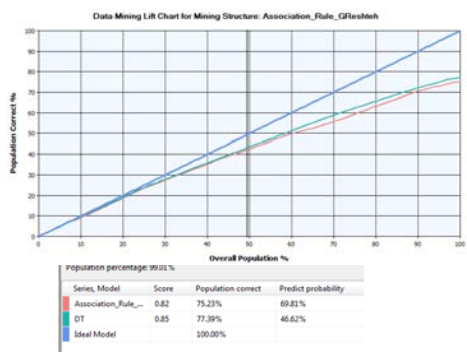

**Figure 6. The Lift chart and  Mining Legend of: DT and Association rule**

To demonstrate this, if we consider predict possibility proportion of each model with correct predicted population, we will see that Association Rules model can correctly predict 75 percent of population with a possibility of 70 percent, while generated model by decision tree can only correctly predict 77 percent of population with a possibility of 47 percent. Therefore, according to yielded results, we will use yielded predict model by Association Rules to predict results.

## 6. Execution and surveying selected model

In this study, by using unsupervised algorithm of Association Rules, we made two different predict models. First model predicts rules of educational fields that each candidate may be accepted in (e.g. IT field[3]), in second model, we predict student status (before admission) of applicants, that is, according to current status of candidate, we extract the rules which predict the candidate's belonging to each of three classes in section 3.

After execution of Association Rules algorithm, in both models, we gained some rules. For instance, some of the generated rules in former model are observed in table through which a model is generated to propose field for candidates of next year. In table 2, field column is the RHS part of the rule.

Figure 7 and 8 show lift chart and Mining Legend table of performance of first and second models. Right hand chart shows how first modes rises and left hand chart shows how second model rises. Red colored curve is lift procedure of the models

---

[3] This university  accepts  students in 4 group fields and more than 160 fields every year

and blue colored curve is ideal model whose predict precision is 100%.

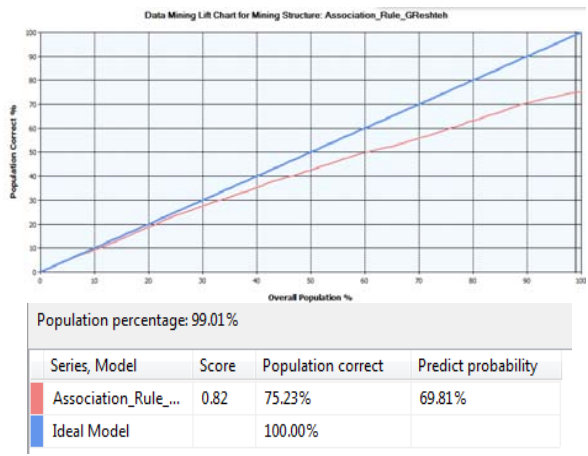

**Figure7.Lift chart and Mining Legend table of performance of first model**

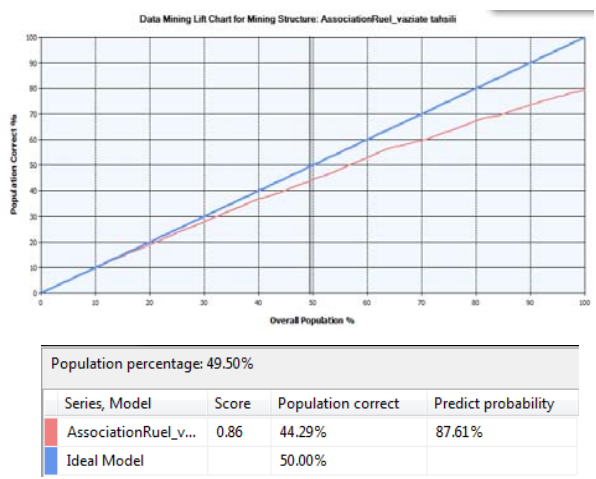

**Figure 8. Lift chart and Mining Legend table of performance of second model**

Yielded results reveal that both predict models present a relatively suitable performance. Therefore, according to yielded results, we applied both Association Rules on datasets of 2009 accepted candidates, so that former model was used to predict accepted field and latter one was utilized to predict student status of applicants.

## 7. Conclusions

According to variety of universities, having such a system can be very effective in order to measure and admit candidates and also encourage and increase scientific level of employees having more than 25 years. Usually, admission of association degree happens in low ages and after graduating in high school but the performed study can prove mission of UAST to improve scientific level of employees. In yielded results by synthetic clustering of modular courses accepted candidates, it can be observed that more than a half of accepted persons are employees and half of them have been educated in a field related to their occupation. Also, by surveying clusters, it can be seen that student status of unemployed candidates is better than other accepted candidates and also they have less age.

Generally in this case study, in addition to data mining on data set, our aim is to select the most suitable algorithms and models; so that, initially to execute clustering algorithm, we considered and assessed existing challenges. Also, by this grouping, we labeled educational dataset with three separate classes and next problem was to select suitable algorithm to predict results and student status of candidates for which Association Rules algorithm was chosen according to performed surveys. By using such an algorithm, we did extraction of rules and generating predict model. Yielded results were approximately compatible with policy of the university. By performing this study, we hope to propose a model for applicants and candidates of such courses so that it would be able to help them make better decisions to select field and specify student status.

## 8. Acknowledgements

This work was financially supported by one of the branch of University of Applied Science and Technology. The authors express the utmost appreciation for their sample collection and also helping, developing and testing this prediction model.